\documentclass[11pt,twoside]{article}
\usepackage{CAGN2016}
\usepackage{graphicx}

\usepackage[T1]{fontenc} 

\usepackage{latexsym}
\usepackage{verbatim}

\begin{document}

\vskip 1.0cm
\markboth{L.~Toribio San Cipriano et al.}{Spatial distribution of C and O in MCs}
\pagestyle{myheadings}
%
%
\vspace*{0.5cm}
\parindent 0pt{Contributed  Paper}


\vspace*{0.5cm}
\title{Spatial distribution of carbon and oxygen abundances in the Magellanic Clouds}

\author{L.~Toribio San Cipriano$^{1,2}$, C.~Esteban$^{1,2}$, G.~Dom\'inguez-Guzm\'an$^{3}$ and J.~Garc\'ia-Rojas$^{1,2}$}
\affil{$^{1}$Instituto de Astrof\'isica de Canarias, E-38200, La Laguna, Tenerife, Spain\\
$^{2}$Departamento de Astrof\'isica, Universidad de La Laguna, E-38206, La Laguna, Tenerife, Spain\\
$^{3}$Instituto Nacional de Astrof\'isica, \'Optica y Electr\'onica, Apartado Postal 51 y 216, Puebla, Mexico\\}

\begin{abstract}

We present chemical abundances of carbon and oxygen in the Large and Small Magellanic Clouds from optical spectra of H\thinspace\textsc{ii} regions.  We analyze the behavior of the O/H, C/H and C/O abundances ratios and their spatial distribution inside the galaxies. The results show that the radial gradients can be considered flat for all these elements in both galaxies. In addition, we compare our results with those of other more massive spiral galaxies. We find a correlation between the absolute magnitude, $M_\mathrm{V}$, of the galaxies and the slopes of C/H and C/O gradients. The more massive galaxies show steeper C/H and C/O gradients than the less massive  ones. 

\end{abstract}

\section{Introduction}

A proper knowledge of the spatial distribution of chemical abundances in galaxies is crucial for building up chemical evolution models. In particular, H\thinspace\textsc{ii} regions are used as the best tracers of the current composition of the extragalactic interstellar medium since they can be observed at large distances. After hydrogen and helium, oxygen (O) and carbon (C) are the third and fourth  most abundant elements in the Universe, respectively. O abundance of H\thinspace\textsc{ii} regions is the most used proxy of the metallicity of galaxies. Despite the importance of C as a source of opacity in stars and its fundamental role in interstellar dust and organic molecules, C abundance in extragalactic  H\thinspace\textsc{ii} regions has been poorly explored. Garnett~et~al. (1995, 1999) derived C abundances in several extragalactic H\thinspace\textsc{ii} regions of nearby spiral galaxies using the C\thinspace\textsc{iii}] 1909 \AA\ and  C\thinspace\textsc{ii}] 2326 \AA\ collisionally excited lines (hereinafter CELs). Recently, Berg~et~al. (2016) reported C abundances from UV CELs for 12 H\thinspace\textsc{ii} regions in nearby dwarf galaxies. However, abundances computed from UV CELs have the disadvantage that they are strongly affected by the uncertainty in the choice of UV reddening function. Alternatively, there is another method to derive C abundances in H\thinspace\textsc{ii} regions. Thanks to the arrival of large aperture telescopes and to the more efficient CCDs, especially in the blue range, reliable measurements of the faint C\thinspace\textsc{ii} 4267 \AA\ recombination line (hereinafter RL) have been achieved. In the last years, several authors (e.g. Peimbert~et~al. 2005; Esteban~et~al. 2002, 2009, 2014; L\'opez-S\'anchez~et~al. 2007; Toribio San Cipriano~et~al. 2016) have determined C abundances in Galactic and extragalactic  H\thinspace\textsc{ii} regions based on this method.

In this work we present C abundances derived from RLs and O ones from RLs and CELs in the Large Magellanic Cloud (LMC) and the Small Magellanic Cloud (SMC) from deep, high-quality optical spectra of H\thinspace\textsc{ii} regions. We explore the distribution of O/H, C/H and C/O ratios and discuss the results in the framework of the chemical evolution of galaxies.


\section{Sample}
\label{sample}

We used a sample of five H\thinspace\textsc{ii} regions in the LMC and four H\thinspace\textsc{ii} regions in the SMC. Seven of these objects were observed at Cerro Paranal Observatory (Chile) with the Ultraviolet Visual Echelle Spectrograph (UVES) mounted at the Kueyen unit of the 8.2-m Very Large Telescope (VLT). The data of the other two H\thinspace\textsc{ii} regions (one in the LMC and other in the SMC) were taken from Peimbert (2003) and Pe\~na-Guerrero et al. (2012). In order to have an homogeneous data set, in these two cases, we took the measured line fluxes and performed the analysis in the same way than the rest of the sample. The description and analysis of this data set can be found in Dom\'inguez-Guzm\'an et al. (this proceedings).

We measured the faint C\thinspace\textsc{ii} 4267 \AA\ RL and all or most of the lines of multiplet 1 of the O\thinspace\textsc{ii} RLs at about 4650 \AA\ in all the objects of the sample, allowing the computation of $\mathrm{C}^{2+}$ and $\mathrm{O}^{2+}$ abundances. In addition, we derived $\mathrm{O}^{2+}$ from CELs as well as the total abundances of C and O (from CELs and RLs). The physical conditions used for the determination of the abundances were defined assuming a two-zone scheme. The low-ionization zone, characterized by the electron temperature $T_\mathrm{e}([\mathrm{N}\thinspace\textsc{ii}])$ and the weighted mean of the electron density $n_\mathrm{e}([\mathrm{S}\thinspace\textsc{ii}])$ and $n_\mathrm{e}([\mathrm{O}\thinspace\textsc{ii}])$, was used to calculate the $\mathrm{O}^{+}$ abundance. The high-ionization zone, characterized by $T_\mathrm{e}([\mathrm{O}\thinspace\textsc{iii}])$ and $n_\mathrm{e}([\mathrm{Cl}\thinspace\textsc{iii}])$, was used to derive the $\mathrm{C}^{2+}$ and $\mathrm{O}^{2+}$ abundances. In order to correct for the unseen ionization stages of the C, we used the ICF by Garnett (1999). In addition, for one of the H\thinspace\textsc{ii} regions which has He\thinspace\textsc{ii} lines in the spectrum, we corrected the $\mathrm{O^{3+}}$ contribution adopting the ICF by Delgado-Inglada et al. (2014).

\section{Results}\label{result}

\begin{figure}  
\begin{center}
\hspace{0.25cm}
\includegraphics[width=0.48\textwidth]{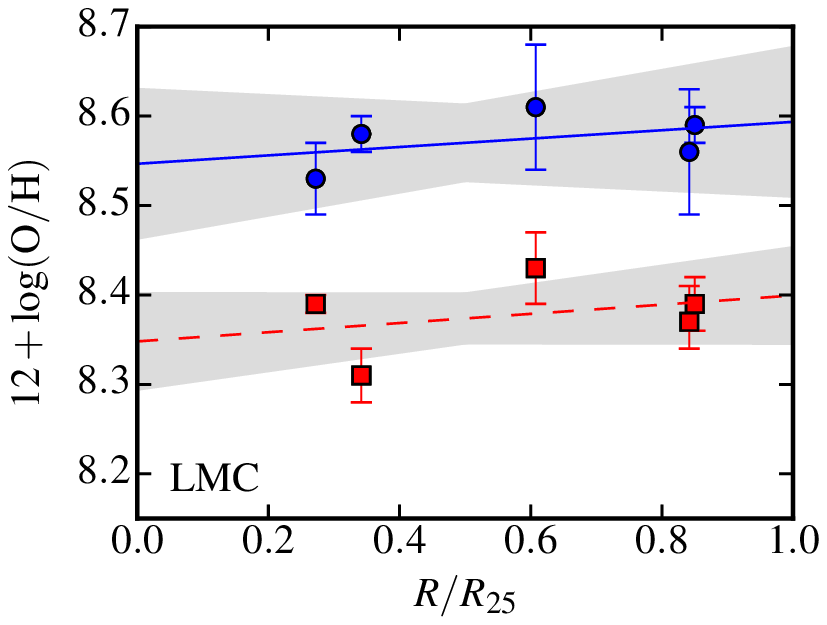}
\includegraphics[width=0.48\textwidth]{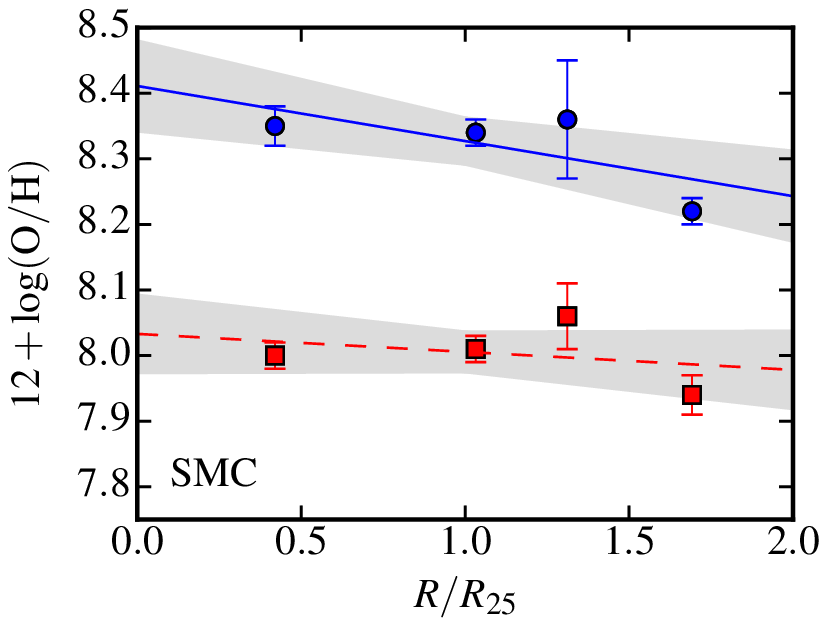}
\caption{Spatial distribution of the O abundances (blue circles from RLs and red squares from CELs) as a function of $R/R_{25}$ for the LMC (left-hand panel) and the SMC (right-hand panel). The grey areas trace all the radial gradients compatible with the uncertainties of our least-squares linear fits computed through Monte Carlo simulations.}
\label{fig:o}
\end{center}
\end{figure}

We studied the spatial distribution of O and C abundances in order to explore the presence of radial abundance gradients in the Magellanic Clouds (MCs). We performed least-squares linear fits to the fractional galactocentric distance of the objects, ($R/R_{25}$)\footnote{Defined as the radius where the B-band surface brightness is 25 mag arcsec$^{-2}$.}, for both galaxies. In the case of radial O abundance gradients, we have two different fits whether $\mathrm{O}^{2+}/\mathrm{H}^+$ was determined from RLs or CELs. For the LMC:
\begin{equation}
\label{eq:o_RLs_LMC}
12 + \log(\mathrm{O/H})_{\mathrm{RLs}} = 8.55 (\pm 0.04) + 0.04 (\pm 0.07)R/R_{25},
\end{equation}
\begin{equation}
\label{eq:o_CELs_LMC}
12 + \log(\mathrm{O/H})_{\mathrm{CELs}} = 8.35 (\pm 0.03) + 0.05 (\pm 0.05)R/R_{25},
\end{equation}
and for the SMC:
\begin{equation}
\label{eq:o_RLs_SMC}
12 + \log(\mathrm{O/H})_{\mathrm{RLs}} = 8.41 (\pm 0.04) - 0.09 (\pm 0.04)R/R_{25},
\end{equation}
\begin{equation}
\label{eq:o_CELs_SMC}
12 + \log(\mathrm{O/H})_{\mathrm{CELs}} = 8.03 (\pm 0.03) - 0.03 (\pm 0.03)R/R_{25}.
\end{equation}
Figure \ref{fig:o} represents the spatial distribution of the O abundances as a function of $R/R_{25}$ and the linear fits for the LMC (left-hand panel) and the SMC (right-hand panel). The blue circles represent the abundances determined from RLs and the red squares those from CELs. The grey areas trace all the radial gradients compatible with the uncertainties of our least-squares linear fits computed through Monte Carlo simulations. The offset between both determinations of O/H ratios (CELs and RLs) for the same object represents the effect known as \textit{abundance discrepancy}. It refers to the difference between the chemical abundances determined from RLs and from CELs of the same ion (see Esteban et al., this proceedings). If we consider the slope of the gradients represented in Figure \ref{fig:o} and defined in equations \ref{eq:o_RLs_LMC} and \ref{eq:o_CELs_LMC} for the LMC and \ref{eq:o_RLs_SMC} and \ref{eq:o_CELs_SMC} for the SMC, we can conclude that for both galaxies the slopes from CELs and RLs agree within the errors. In previous works for other galaxies, Esteban et al. (2005, 2009) and Toribio San Cipriano et al. (2016) also reported similar slopes of the O/H gradients independently of the kinds of lines used to derive the abundances. The O/H gradients obtained can be considered rather flat in both galaxies, taking into account the uncertainties computed for the slopes.

\begin{figure}  
\begin{center}
\hspace{0.25cm}
\includegraphics[width=0.48\textwidth]{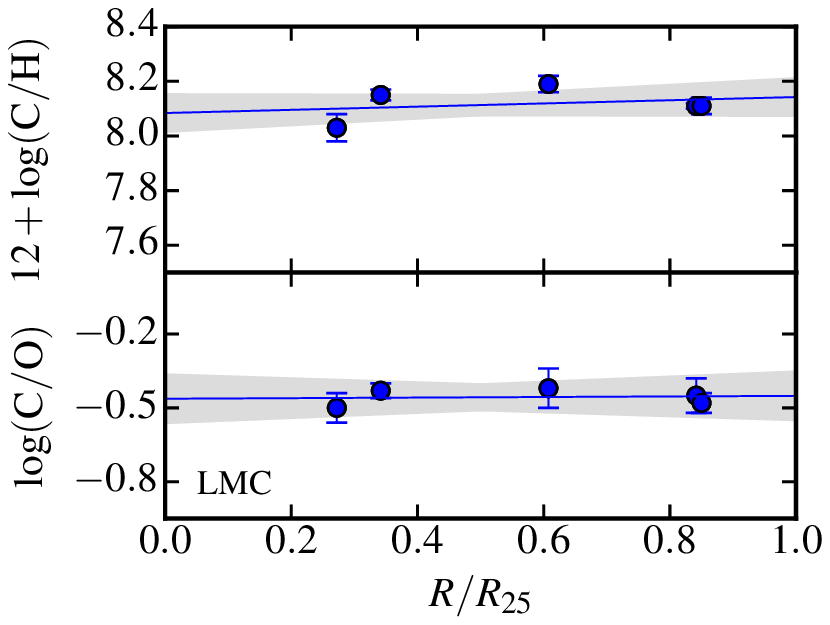}
\includegraphics[width=0.48\textwidth]{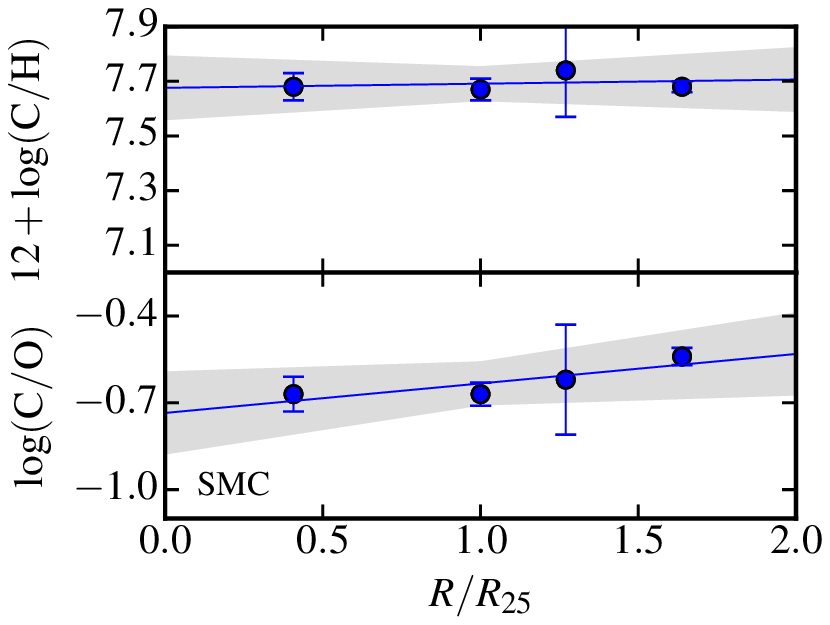}
\caption{Spatial distribution of the C/H (upper panels) and C/O (lower panels) abundances as a function of the fractional galactocentric distance, $R/R_{25}$, for the LMC (left-hand panel) and the SMC (right-hand panel). The grey areas trace all the radial gradients compatible with the uncertainties of our least-squares linear fits computed through Monte Carlo simulations.}
\label{fig:c}
\end{center}
\end{figure}

Figure \ref{fig:c} represents the spatial distribution of the C/H (upper panels) and C/O (lower panels) ratios as a function of $R/R_{25}$ for both galaxies (LMC left-hand panel and the SMC right-hand panel). The least-squares linear fits for the LMC are:
\begin{equation}
\label{eq:c_LMC}
12 + \log(\mathrm{C/H}) = 8.08 (\pm 0.05) + 0.06 (\pm 0.07)R/R_{25},
\end{equation}
\begin{equation}
\label{eq:cvs_LMC}
\log(\mathrm{C/O}) = -0.46 (\pm 0.06) + 0.01 (\pm 0.10)R/R_{25},
\end{equation}
and for the SMC:
\begin{equation}
\label{eq:c_SMC}
12 + \log(\mathrm{C/H}) = 7.68 (\pm 0.06) + 0.01 (\pm 0.06)R/R_{25},
\end{equation}
\begin{equation}
\label{eq:cvs_SMC}
\log(\mathrm{C/O}) = -0.73 (\pm 0.07) + 0.09 (\pm 0.07)R/R_{25}.
\end{equation}
We can conclude that the slope of C/H and C/O in both galaxies can be considered practically flat, in contrast with the results we have found in spiral galaxies where the gradients are negative. 

Other authors also reported flat radial gradients in the MCs. Pagel et al. (1978) studied the spatial distribution of O abundance in H\thinspace\textsc{ii} regions of both galaxies finding small or flat gradients. Cioni (2009) reported a smooth gradient of the Fe/H ratio in the LMC and a negligible one in the SMC from an asymptotic giant brach (AGB) stars. Recently, Choudhury et al. (2016) found a similar slope than ours from a study of red giant brach (RGB) stars in the LMC.

\begin{figure}  
\begin{center}
\hspace{0.25cm}
\includegraphics[height=5.0cm]{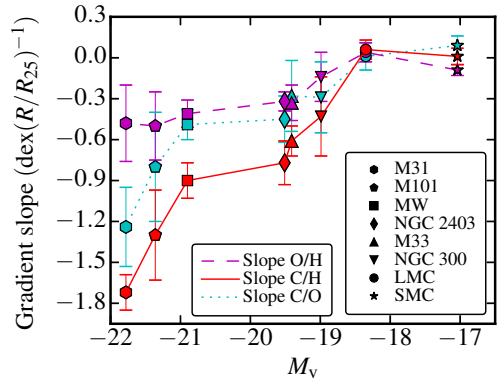}
\caption{Slope of the O/H (magenta dashed line/symbols), C/H (red continuous line/symbols), and C/O (cyan dotted line/symbols) radial gradients versus absolute magnitude, $M_\mathrm{V}$ of several spiral galaxies. Figure taken from Toribio San Cipriano et al. (2017).}
\label{fig:relation}
\end{center}
\end{figure}

Figure \ref{fig:relation} represents the slope of O/H (magenta dashed line/symbols), C/H (red continuous line/symbols) and C/O (cyan dotted line/symbols) radial gradients with respect to the absolute magnitude, $M_\mathrm{V}$, for several galaxies. This figure reinforces the results obtained by Toribio San Cipriano (2016) where the authors found that the slopes of C/H and C/O gradients show a correlation with  $M_\mathrm{V}$ of the galaxies. The more luminous galaxies as M\,31 or M\,101 show systematically steeper C/H and C/O gradients than the less luminous ones (e.g. NGC\,300 or MCs). According with the results of Figure \ref{fig:relation}, the mechanism of C enrichment would be more effective in the more massive galaxies than in the less massive ones. A deeper analysis of this figure is presented in Toribio San Cipriano et al. (2017).

\section{Conclusions}
We present the spatial distribution of O/H, C/H and C/O ratios in the Magellanic Clouds through a study of the C\thinspace\textsc{ii} and O\thinspace\textsc{ii} recombination lines in H\thinspace\textsc{ii} regions. We find that radial gradients are practically flat within the uncertainties in both galaxies. Other authors also report flat gradients in the MCs from RGB and AGB stars. We compare our results with those in other galaxies finding that the gradients seem to follow the trend described by Toribio San Cipriano et al. (2016) where the more massive galaxies show steeper C and C/O gradients than the less luminous galaxies.

\acknowledgments This work is based on observations collected at the European Southern Observatory, Chile, proposal numbers ESO 092.C-0191(A) and ESO 60.A-9022(A). LTSC is supported by the FPI Program by the Ministerio de Econom\'{i}a y Competitividad
(MINECO) under grant AYA2011-22614. This project is also partially funded by MINECO under grant AYA2015-65205-P.

\end{document}